\documentclass[a4paper,11pt]{article}
\pdfoutput=1
\usepackage{jheppub}
\usepackage{amsmath}
\usepackage{slashed}
\usepackage{amssymb}
\usepackage{color}
\usepackage{graphicx}
\usepackage{hyperref}
\usepackage{xspace,slashed}
\usepackage[utf8]{inputenc}
\usepackage{subcaption}
\usepackage{multirow}
\usepackage{algorithm}
\usepackage{algorithmic}
\usepackage{stackengine}
\usepackage{enumitem}
\usepackage{etoolbox}

\newtoggle{draft}

\togglefalse{draft}

\iftoggle{draft} {
  \usepackage{changes}
}{%
  \usepackage[final]{changes}
}

\definechangesauthor[name=pn, color=blue]{pn}
\definechangesauthor[name=km, color=magenta]{km}

\usepackage{booktabs}

\makeatletter
\g@addto@macro\bfseries{\boldmath}
\makeatother

\newcommand{\bp}{{\bf p}}
\newcommand{\bk}{{\bf k}}

\newcommand{\be}{\begin{equation}}
\newcommand{\ee}{\end{equation}}

\newcommand{\bnq}{\begin{equation}}
\newcommand{\enq}{\end{equation}}
\newenvironment{eqaligned}{\equation\aligned}{\endaligned\endequation}
\newcommand{\bns}{\begin{eqaligned}}
\newcommand{\ens}{\end{eqaligned}}

\newcommand{\nn}{\nonumber \\}

\newcommand{\TTPaff}{Institute for Theoretical Particle Physics,
  KIT, 76128 Karlsruhe, Germany}

\allowdisplaybreaks

\preprint{
    TTP23-036, P3H-23-060 
}

\title{
  Scale dependence of non-factorizable virtual corrections
  to Higgs boson production in weak boson fusion
}
\author[]{Christian Br\o{}nnum-Hansen,}
\author[]{Ming-Ming Long,}
\author[]{Kirill Melnikov}

\affiliation[]{\TTPaff}

\emailAdd{christian.broennum-hansen@partner.kit.edu}
\emailAdd{ming-ming.long@kit.edu}
\emailAdd{kirill.melnikov@kit.edu}

\abstract{
The renormalization-scale  dependence of 
  the non-factorizable virtual corrections to Higgs boson production in weak boson
  fusion at next-to-next-to-leading order in perturbative  QCD is unusually strong, due to the  peculiar nature of these
  corrections.
  To address this problem, we  compute the three-loop non-factorizable contribution to  this process which  
  accounts for the running of the strong
  coupling constant,   and show that it stabilizes the theoretical prediction.
    }

\begin{document}

\maketitle 
\section{Introduction}

The non-factorizable corrections
to Higgs boson production in weak boson fusion (WBF) are color-suppressed
and, for this reason,
are expected to be smaller than the factorizable
ones.~\footnote{Theoretical predictions for Higgs boson
  production in WBF are very advanced, see Refs.~\cite{Figy:2003nv,Berger:2004pca,Ciccolini:2007ec,Andersen:2007mp,Harlander:2008xn, Bolzoni:2010xr,Bolzoni:2011cu,Cacciari:2015jma,Dreyer:2016oyx}, for the discussion of factorizable
  QCD and electroweak corrections,  as well as effects of multi-jet merger and
  interplay of parton showers and fixed-order predictions \cite{Chen:2021phj}.}
However, virtual non-factorizable corrections,  which
start contributing to the WBF cross section  at next-to-next-to-leading order (NNLO) in perturbative
QCD,   exhibit a peculiar
enhancement by two powers of $\pi$. This enhancement was first observed when the two-loop amplitude 
was computed in the leading eikonal approximation in Ref.~~\cite{Liu:2019tuy}. Recently, the calculation
reported in Ref.~\cite{Liu:2019tuy} was extended in two important ways. First,
in Ref.~\cite{Asteriadis:2023nyl} the calculation of real-virtual and double-real non-factorizable
corrections to Higgs boson production in weak boson fusion was performed.  It was shown that these contributions
are negligible in comparison with ${\cal O}(\pi^2)$-enhanced eikonal virtual corrections. Second, in Ref.~\cite{Long:2023mvc}
it was found that the leading power correction to the eikonal contribution 
reduces the prediction based on
the eikonal approximation by about $20$ percent.

An important observation that emerged from the studies described above is that the dependence of
these predictions on the renormalization scale is significant and can easily reach $20-30$ percent.\footnote{To avoid confusion, we emphasize that we refer to  the non-factorizable
  contribution only.} This feature is the direct consequence
of the fact that the non-factorizable corrections appear \emph{for the very first time} at  NNLO
in the perturbative QCD expansion. For this reason
the mechanism responsible for compensating the dependences of physical quantities on the  renormalization scale
is  not yet  present in the results of 
Refs.~\cite{Liu:2019tuy, Asteriadis:2023nyl, Long:2023mvc}, 
in spite of the fact that they are part of the NNLO QCD corrections to Higgs boson production in WBF.

To address this problem,  one should calculate the ${\rm N}^3{\rm LO}$ QCD non-factorizable corrections. This calculation
is currently unfeasible. A much simpler  possibility  is to compute the corrections associated with the running of the
QCD coupling constant, 
starting from the fermion bubble insertions into the gluon  propagators, and then re-writing such  contributions through the full
QCD $\beta$-function in the spirit of the BLM approach \cite{Brodsky:1982gc}. Since, as follows from
Ref.~\cite{Liu:2019tuy}, the leading contribution to the non-factorizable corrections to WBF is related to the potential
scattering, it appears quite natural to include an improved description of the interaction potential between two quarks in QCD as
the first step towards a more reliable theoretical prediction.  This is what we do in this paper.

The rest of the paper is organized as follows. In Section~\ref{sect:bubble} we explain how to accommodate  the running
coupling constant into the computation of non-factorizable corrections and 
derive a convenient one-dimensional representation for the  three-loop ${\cal O}(\beta_0 \alpha_s^3)$
corrections. In Section~\ref{sect:analytic} we discuss the  analytic computation of these contributions.
In Section~\ref{sect:numres} we show
that the computed corrections  strongly reduce the renormalization scale dependence of the theoretical predictions for
the  non-factorizable contributions to Higgs boson production in
WBF,  making them more precise and reliable.  We conclude in Section~\ref{sect:conclusion}.   Analytic formulas for ${\cal O}(\beta_0 \alpha_s^3)$
corrections to Higgs boson production in WBF can be found in an ancillary file provided with this submission.

\section{Fermion-bubble corrections}
\label{sect:bubble}

\begin{figure}[h]
\centering
\includegraphics[width=0.2 \textwidth]{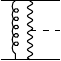} 
	\hspace{2cm}
\includegraphics[width=0.2 \textwidth]{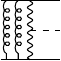}
	\caption{Representative Feynman diagrams that contribute to the NNLO non-factorizable corrections to Higgs production in WBF.}
\label{fig: feynman diagram 1}
\end{figure}

According to the analyses of Refs.~\cite{Liu:2019tuy,Long:2023mvc}
the leading ${\cal O}(\alpha_s^2)$ QCD non-factorizable contributions to Higgs boson production in weak boson fusion 
can be written in the following way
\be
   {\rm d} \sigma_{\rm nf} =
   \frac{N_c^2-1}{4 N_c^2} \alpha_s^2 \;  {\rm d} \sigma^{\rm LO} \; C_{\rm nf}\,.
   \label{eq2.1}
\ee
Here $N_c=3$ is the number of colors, $\alpha_s$ is the strong coupling constant and
${\rm d} \sigma^{\rm LO}$ is the Born differential cross section.
Eq. (\ref{eq2.1}) is obtained by computing the square of the one-loop amplitude and the
interference of the two-loop and Born amplitudes in the eikonal approximation,
see Figure \ref{fig: feynman diagram 1} for representative Feynman diagrams.

The function $C_{\rm nf}$, which describes the effect of the non-factorizable corrections, reads 
\be
C_{\rm nf} = C_1^2 - C_2, 
\ee
where  
\be
\begin{split}
  & C_1 = -2 \int \frac{ {\rm d}^{d-2} \bk_{1}}{ (2 \pi)^{d-3} }
\frac{\Delta_3 \Delta_4}{ \Delta_1 \Delta_{3,1} \Delta_{4,1}},
\\
&
C_2 = 4 \int \frac{ {\rm d}^{d-2} \bk_{1}}{ (2 \pi)^{d-3} } \frac{ {\rm d}^{d-2} \bk_{2}}{ (2 \pi)^{d-3} }
\frac{\Delta_3 \Delta_4}{ \Delta_1 \Delta_2 \Delta_{3,12} \Delta_{4,12}}.
\end{split}
\ee
In the above equation, we have used 
\be
\Delta_i = \bk_i^2,\;\;\; \Delta_{3,i} = (\bk_i - \bp_{3} )^2 + m_V^2,
\;\;\;\;\Delta_{4,i} = (\bk_i + \bp_{4} )^2 + m_V^2, \quad i=1,2,12,
\ee
as well as $\bk_{12} = \bk_1 + \bk_2$ and 
\be
\Delta_3 = \bp_3^2 + m_V^2,\;\;\; \Delta_4 = \bp_4^2 + m_V^2.
\ee
Furthermore, $m_V$ is the mass of the electroweak gauge boson\footnote{Note that in Eq.~(\ref{eq2.1})
one needs to sum over contributions of $Z$ and $W$ bosons. } and we employ boldface fonts to denote two-dimensional Euclidean vectors.
For example,  $\bp_{3,4}$ are the transverse momenta of the tagging jets in the WBF process $pp \to 2j + H$.
We also use $\bp_H$ to denote the transverse momentum of Higgs.

\vspace*{0.3cm}
An important feature of non-factorizable corrections is that the
function $C_{\rm nf}$ is infrared finite 
although the functions $C_{1,2}$ are infrared divergent.   To see this, it is simplest
to use the integral representations for $C_{1,2}$ and write $C_{\rm nf}$ as 
\be
C_{\rm nf} = 
4 \int \frac{ {\rm d}^{d-2} \bk_{1}}{ (2 \pi)^{d-3} } \frac{ {\rm d}^{d-2} \bk_{2}}{ (2 \pi)^{d-3} }
\frac{\Delta_3 \Delta_4}{ \Delta_1 \Delta _2}
\left (
\frac{\Delta_3 \Delta_4} { \Delta_{3,1} \Delta_{4,1} \Delta_{3,2} \Delta_{4,2}}
- \frac{1}{\Delta_{3,12} \Delta_{4,12}} \right ).
\label{eq2.5}
\ee
Using explicit expressions for the $\Delta$-functions, it is easy to  show that
\be
\frac{\Delta_3 \Delta_4} { \Delta_{3,1} \Delta_{4,1} \Delta_{3,2} \Delta_{4,2}}
- \frac{1}{\Delta_{3,12} \Delta_{4,12}}
= \bk^i_{1} \bk^j_{2} \; T_{ij}(\bk_{1}, \bk_{2},..),
\ee
where the rank-2 tensor $T_{ij}$ is non-singular  in the limit of vanishing $\bk_{1,2}$.
This fact ensures that $C_{\rm nf}$ is not infrared divergent  and, as a consequence,
in Eq.~(\ref{eq2.5}) we can replace the space-time dimensionality $d$ with $4$, so that $d-2 \to 2$. 

\begin{figure}[h]
\centering
\includegraphics[width=0.2 \textwidth]{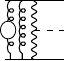} 
	\caption{A representative fermion-bubble insertion diagram that contributes to the $\mathcal{O}(\beta_0 \alpha_s^3)$ corrections. One should sum over $N_f$ flavors running in the bubble.}
\label{fig: feynman diagram 2}
\end{figure}

It is now straightforward to introduce the running coupling constant
into Eq.~(\ref{eq2.5}) since all  we need to do is to 
modify the gluon propagators $\Delta_1$ and $\Delta_2$ in  Eq.~(\ref{eq2.5}). We find\footnote{For simplicity, we still use $C_{\text{nf}}$ instead of introducing a new notation.}
\be
C_{\rm nf} = 
4 \int \frac{ {\rm d}^2 \bk_{1}}{ (2 \pi) } \frac{ {\rm d}^2 \bk_{2}}{ (2 \pi) }
\frac{\Delta_3 \Delta_4}{ {\tilde \Delta}_1 {\tilde \Delta}_2}
\left (
\frac{\Delta_3 \Delta_4} { \Delta_{3,1} \Delta_{4,1} \Delta_{3,2} \Delta_{4,2}}
- \frac{1}{\Delta_{3,12} \Delta_{4,12}} \right ),
\label{eq2.8}
\ee
where
\be
\tilde \Delta_i = \Delta_i \;  \left ( 1 + \frac{\beta_0 \alpha_s}{2 \pi} \ln \frac{\bk_{i}^2}{\mu^2 e^{5/3}}  \right ),
\label{eq2.9}
\ee
$\beta_0 = 11/6C_A - 2/3 N_f T_R$, $C_A =3$, $T_R = 1/2$, $N_f$ is the number
of massless quark flavors and $\mu$ is the renormalization scale of the coupling $\alpha_s$ in Eq.~(\ref{eq2.1}). We note
that by using Eq.~(\ref{eq2.9})  in Eq.~(\ref{eq2.8})
we describe the all-order impact of the running coupling constant
on $C_{\rm nf}$ but, since we are interested in computing ${\cal O}(\beta_0 \alpha_s^3)$ corrections
to Higgs boson production in WBF,  we only use this formula
for bookkeeping purposes. As we explain below,  eventually, we  expand Eq.~(\ref{eq2.8}) 
in powers of $\alpha_s$ to the required order. 
To compute  ${\cal O}(\beta_0 \alpha_s^3)$ corrections, 
we focus on contributions of diagrams with fermion bubbles inserted into 
one of the two gluon propagators, see Figure \ref{fig: feynman diagram 2} for an example.

 Although one can work with $C_{\rm nf}$ in Eq.~(\ref{eq2.8}),
 it is more convenient to compute $C_1$ and $C_2$ separately.  However,
$C_{1,2}$ are infrared divergent so that extracting them from Eq.~(\ref{eq2.8}) and calculating
them separately requires an  infrared regulator. Since this regulator should work efficiently
for integrands with $1/\bk^2$ and $1/\bk^2 \, \ln \bk^2$, it seems that the best option is an \emph{analytic regulator}
where each gluon propagator is raised to the power $1+ \nu$.  
Hence, we define 
\be
\begin{split}
  & C_{1}(\nu) = -2 \int \frac{ {\rm d}^2 \bk_{1}}{ 2 \pi } \; 
\frac{\Delta_3 \; \Delta_4 \; m_V^{2\nu} }{ \Delta_1^{1+\nu} \Delta_{3,1}  \Delta_{4,1}},
\\
&
C_{2}(\nu_1, \nu_2)= 4 \int \frac{ {\rm d}^2 \bk_{1}}{ 2 \pi } \frac{ {\rm d}^2 \bk_{2}}{ 2 \pi }
\frac{\Delta_3 \; \Delta_4 \; m_{V}^{2 ( \nu_1 + \nu_2) }} { \Delta_1^{1+\nu_1} \Delta_2^{1+\nu_2}
 \Delta_{3,12} \Delta_{4,12}},
\end{split}
\label{eq2.10}
\ee
and we will use these two auxiliary functions to compute $C_{\rm nf}$ through ${\cal O}(\beta_0 \alpha_s)$.

We begin with the calculation of $C_{1}(\nu)$. Introducing Feynman parameters and integrating over the transverse momentum,
we obtain the following representation for this function 
\be
\begin{split} 
	C_{1}(\nu) &= - \frac{\Gamma(2 + \nu) \Delta_x \Delta_y  }{\Gamma(1+ \nu)} 
\int \limits_{0}^{1} {\rm d} t \; \int \limits_{0}^{1} \;
\frac{{\rm d} \xi \; \xi^{-1-\nu} ( 1- \xi)^\nu }{ \left (r_{12}(t) - r_1(t) \; \xi   \right )^{2+\nu} }
\\
& = \frac{\Gamma(2+\nu) \Gamma(1-\nu) \; \Delta_x \Delta_y  }{\nu} \int \limits_{0}^{1}  \frac{ {\rm d} t}{  
\left ( r_{12}(t) \right )^{2+\nu} } \; F_{21} \left ( 2+\nu, -\nu, 1, \frac{r_1(t)}{r_{12}(t)} \right ),
\end{split}
\label{eq2.11}
\ee
where $F_{21}$ is the  hypergeometric function.
To write  Eq.~(\ref{eq2.11}) we introduced the following functions
\bnq
\begin{gathered}
 \Delta_x = 1+x,\;\;\;\; \Delta_y = 1+y, \\
 r_1 = x t + y (1-t) - z t(1-t), \;\;\; r_2  = 1 + z t(1-t),\;\;\;\;  r_{12}= r_1 + r_2, 
\end{gathered}
\enq
and the following dimensionless quantities 
\bnq
x = \frac{\bp_{3}^2}{m_V^2}, \quad
y = \frac{\bp_{4}^2}{m_V^2}, \quad
z = \frac{\bp_{H}^2}{m_V^2},
\enq
that we will use throughout the calculation. 

With this result at hand, it is straightforward to compute $C_2(\nu_1, \nu_2)$. Indeed, in this  case
we can first integrate over one of the two loop momenta,  keeping their \emph{sum} fixed.
We obtain 
\be
\begin{split}
& \int \frac{ {\rm d}^2 \bk_{1} }{(2\pi) } 
\frac{1}{( \bk_{1}^2 ) ^{1+\nu_1} (  ( \bk_{12} - \bk_{1})^2 )^{1+\nu_2}}
\\
& =
-
\frac{\nu_{12}}{2 \nu_1 \nu_2} \frac{\Gamma(1+\nu_{12}) }{\Gamma(1+ \nu_1) \Gamma(1 + \nu_2)}
\frac{\Gamma(1-\nu_1) \Gamma(1-\nu_2)}{\Gamma(1 - \nu_{12} ) } \frac{1}{ ( \bk_{12}^2 )^{1+\nu_{12}}
   }, 
\end{split} 
\ee
where $\nu_{12} = \nu_1 + \nu_2$.  Using this expression in Eq.~(\ref{eq2.10}), we find that 
the remaining integration over $\bk_{12}$ is identical to the one-loop case provided that we replace $\nu$ with $\nu_{12}$. 
Hence, we find
\be
C_2(\nu_1, \nu_2) 
= 
\frac{\nu_{12}}{\nu_1 \nu_2}
\frac{\Gamma(1+\nu_{12}) }{\Gamma(1+ \nu_1) \Gamma(1 + \nu_2)}
\frac{\Gamma(1-\nu_1) \Gamma(1-\nu_2)}{\Gamma(1 - \nu_{12} ) }
C_{1}(\nu_{12} ).
\ee

It is now easy to see that to  compute $C_{\rm nf}$ through ${\cal O}(\beta_0\alpha_s)$ we need to  take $\nu_1 = \nu_2 = \nu$, expand
the quantity 
\be
\left ( \frac{\mu^2 e^{5/3}}{m_V^2} \right )^{2\,\nu}  \left ( C_1(\nu)^2  - C_2(\nu, \nu) \right )
\ee
through first order in $\nu$ and
identify $\nu$ with $\alpha_s \beta_0/(2 \pi)$. 
Writing the expansion of   $C_1(\nu)$ in powers of $\nu$ as follows\footnote{Using
Eq.~(\ref{eq2.11}) and the fact that $F_{21}(2+\nu,-\nu,1,x) = 1 +{\cal O}(\nu)$,   it  is easy to check that 
  the first term in such an expansion is $1/\nu$.}  
\be
C_{1}(\nu) = \frac{1}{\nu} + \sum \limits_{i=1}^{} \; C_{1}^{(i)} \nu^i,
\ee
we obtain 
\be
C_{\rm nf}  = C_{\rm nf}^{(0)} +
\frac{\alpha_s \beta_0 }{\pi} \left ( C_{\rm nf}^{(0)}  \ln \left ( \frac{\mu^2 e^{5/3}}{m_V^2} \right )
+ C_{\rm nf}^{(1)}
\right )
+ {\cal O}( \alpha_s^2\beta_0^2 ),
\ee
where
\be
\begin{split} 
& C_{\rm nf}^{(0)} =  \left ( C_{1}^{(0)}\right )^2 - 2 C_{1}^{(1)},\;\;\;\;\; C_{\rm nf}^{(1)} = C_{1}^{(0)}  C_{1}^{(1)} - 3 C_{1}^{(2)} + 2 \zeta_3 .
\end{split}
\ee

We can easily derive convenient one-dimensional integral representations for  the coefficients
$C_1^{(i)}$, $i=0,1,2$,
  by expanding Eq.~(\ref{eq2.11}) in powers of $\nu$~\cite{Huber:2005yg, Huber:2007dx}. We  find 
\be
\begin{split} 
  C_{1}^{(0)}
  & = \int \limits_{0}^{1} {\rm d}t \;  \frac{ \Delta_x \Delta_y }{r_{12}^2}
\left [ \ln r_2  - 2 \ln r_{12}  + \frac{r_2 - r_1}{r_2}  
     \right ],
   \\
 C_{1}^{(1)} & = \int \limits_{0}^{1} {\rm d} t \;  \frac{ \Delta_x \Delta_4 }{r_{12}^2}
\Bigg  [ \frac{1}{2} \ln^2 r_{12}  - \ln r_{12}  \left ( \frac{r_2 - r_1}{r_2}
  + \ln \frac{r_2}{r_{12}}  \right )
\\
  & +2 \ln \frac{r_2}{r_{12}} + \frac{\pi^2}{6}
-{\rm Li}_2\left( \frac{r_1}{r_{12}} \right ) \Bigg ],
\\
C_{1}^{(2)} & =  \int \limits_{0}^{1} {\rm } {\rm d}t  \;  \frac{ \Delta_x \Delta_4 }{r_{12}^2}
\Bigg [
  -\frac{1}{6} \ln^3 r_{12} 
   +   \frac{1}{2} \ln^2 r_{12}  \left ( \frac{r_2 - r_1}{ r_2} + \ln \frac{r_2}{r_{12}}  \right )
\\
  &  +\frac{\pi^2}{6} \frac{r_2-r_1}{r_2}
 +  \ln ^2 \left ( \frac{r_2}{r_{12}} \right )  \ln \frac{r_1}{r_{12}} - 
 \ln r_{12}  \left ( \frac{\pi^2}{6}  + 2 \ln \frac{ r_2}{r_{12}}  - 
     {\rm Li}_2\left ( \frac{r_1}{r_{12}} \right ) \right )
\\
     & -\frac{r_2-r_1}{r_2} {\rm Li}_2\left ( \frac{r_1}{r_{12}} \right )
- \ln \frac{r_2}{r_{12}}  \left ( \frac{\pi^2}{6} - {\rm Li}_2\left(\frac{r_1}{r_{12}} \right ) \right )
+ 2 {\rm Li}_3 \left ( \frac{r_2}{r_{12}} \right ) - 2 \zeta_3
\Bigg ].
\end{split} 
\label{eq2.19}
\ee
We will discuss below how to use these representations to complete the  analytic computation of $C_1^{(0),(1),(2)}$ but 
  we emphasize that the  integral representations in Eq.~(\ref{eq2.19})
  are quite convenient for numerical calculations.

  Before we discuss the analytic computation in full generality,  we study   $C_{\rm nf}$  in
  some kinematic limits where compact formulas can be derived.  
 The simplest case to consider  is when all  transverse momenta are small
 $|\bp_3| \sim |\bp_4| \ll m_V$. Then,  $r_1 \to 0$ and $r_{12} \to 1$. As the result,
 the expression for $C_1(\nu)$
simplifies and we obtain
\be
\lim_{|\bp_{3,4}| \ll m_V} \; C_1(\nu) \approx \frac{\Gamma(2+\nu) \Gamma(1-\nu)}{\nu}.
\ee
We then find
\be
C_{\rm nf}  = 1 - \frac{\pi^2}{3} +
\frac{\alpha_s \beta_0 }{\pi} \left [ \left (  1 - \frac{\pi^2}{3} \right )  \ln \left ( \frac{\mu^2 e^{5/3}}{m_V^2} \right )
-\frac{ \pi^2}{3} + 2\zeta_3 
\right ]
+ {\cal O}( \alpha_s^2\beta_0^2).
\ee
Following Ref.~\cite{Brodsky:1982gc},  we define the appropriate renormalization
scale $\mu_*$ as the scale for which corrections proportional to $\beta_0$ vanish. 
From the above equation it follows that at small transverse momenta, the appropriate renormalization
scale for the non-factorizable corrections in WBF is 
\be
\mu_* = m_V e^{-5/6}\;  e^{\frac{ \pi^2 -6 \zeta_3}{2(3 - \pi^2)} } \approx 0.36~m_V.
\ee

\vspace*{0.3cm}
Another interesting limit is that of the small Higgs boson momentum
$|\bp_H| \ll |\bp_3| \sim |\bp_4|$. Then, $|\bp_3| = |\bp_4|$ and we find 
\be
r_1 \approx x,\;\;\;\; r_{12} = 1+x.
\ee
As the result, the dependence on $t$ in the integrand in  Eq.~(\ref{eq2.11}) disappears and we obtain 
\be
C_1(\nu) \approx \frac{\Gamma(2+\nu) \Gamma(1-\nu) \;  }{\nu}
 \Delta_x^{-\nu}
 \; F_{21} \left ( 2+\nu, -\nu, 1, \frac{x}{x+1}  \right ).
\ee
It is then straightforward to compute $C_{\rm nf}$ also in this case by expanding the hypergeometric function in
powers of $\nu$. We do not provide the result of such a calculation here and do not discuss the corresponding ``optimal'' scale
choice because  $C_{\rm nf}$ in this case is not positive definite,  which  leads to pathological results
for $\mu_*$. However, we note that we can investigate the leading term in the expansion of
$C_{\rm nf}$ in the limit $|\bp_3| \gg m_V$.
In this case, the leading asymptotic behavior is described by the following formula

\be
C_{\rm nf} \approx x^2 \left [ 1 + \frac{\beta_0 \alpha_s}{ \pi}
\left (\ln \frac{\mu^2 e^{5/3}}{m_V^2}  -  \ln x  \right ) 
+ {\cal O}\left(\alpha_s^2\beta_0^2\right)
\right ].
\ee
 It follows that the appropriate renormalization scale in this case
is
\be
\mu_* = |\bp_3|  e^{-5/6} \approx 0.4 \; |\bp_3|.
\label{eq2.26}
\ee

In practice, the  transverse momenta of the outgoing jets relevant for
Higgs production in weak boson
fusion are  below  $200~{\rm GeV}$. For such transverse momenta,  the formula for the
renormalization scale in  Eq.~(\ref{eq2.26})  leads to  $\mu_* \le m_V \sim m_H$. Therefore, since 
for the smallest  transverse momenta  $\mu_* \sim 0.4 m_V$ and  for the highest relevant
transverse momentum  $\mu_* \sim m_H $, it appears that
the traditional
choice of the scale variation interval used in fixed-order computations of Higgs production in WBF,
roughly covers the two extreme choices of the renormalization scale discussed above. 
In Section~\ref{sect:numres}  we  will illustrate 
that this is indeed the case  by an explicit numerical computation  for the fiducial cross section
and many kinematic distributions.

\section{Analytic computation of the function $C_1$}
\label{sect:analytic}

In this section, we discuss the analytic computation of the function $C_1(\nu)$. 
In principle, such a computation requires a straightforward application of partial fractioning 
and integration by parts. However, one should perform these operations in the right order to minimize the
size of intermediate expressions to the extent possible. Furthermore, it is important
to use symmetries to write  the result as compactly  as possible. To this end, we note that the
expression for $C_1(\nu)$ in
Eq.~(\ref{eq2.11}) is invariant under the interchange of $x$ and $y$,
 and we will use
this symmetry when writing the result of the integration. 

We find it convenient to 
integrate by parts in Eq.~(\ref{eq2.11}) first. 
We obtain 
\bns
C_1(\nu) 
= I_1(\nu)  + I_2(\nu),
\ens
where
\bns
\label{eq3.5}
I_1 &= -\frac{\Gamma(2+\nu)\Gamma(1-\nu) \Delta_x \Delta_y}{\nu(1+\nu)(x-y)} \;\;
	\frac{F_{21}\left(2+\nu,-\nu,1,\frac{r_1(t)}{r_{12}(t)}\right)}{[r_{12}(t)]^{1+\nu}}
	\; \Bigg|_0^1 \;
          , \\
I_2 &= \frac{\Gamma(3+\nu)\Gamma(1-\nu) \Delta_x \Delta_y}{(1+\nu)(y-x)}
\int \limits_{0}^{1}
\frac{{\rm d}t}{[r_{12}(t)]^{1+\nu}}F_{21}\left(3+\nu,1-\nu,2,
\frac{r_1(t)}{r_{12}(t)}\right) \frac{\rm d}{{\rm d}t} \frac{r_1(t)}{r_{12}(t)}.
\ens

It is straightforward to compute $I_1$. We find 
\bnq
I_1 = \frac{\Gamma (1-\nu ) \Gamma (\nu +2) (x+1) (y+1)^{-2 \nu } \, F_{21}(-\nu -1,-\nu ;1;-y)}{\nu  (\nu
   +1) (x-y)} + (x \leftrightarrow y),
\enq
where  the symmetry under the interchange of $x$ and $y$ is made manifest.
Expanding the above formula  in powers of $\nu$, we obtain 
\bnq
I_1 = \frac{1}{\nu} + I_1^{(0)} + I_1^{(1)} \, \nu + I_1^{(2)} \, \nu^2 + \mathcal{O}(\nu^3),
\enq
where
\bns
I_1^{(0)} &= -\frac{x+1}{x-y}\left[y+2 \ln (y+1)\right]+ (x \leftrightarrow y), \\
I_1^{(1)} &= 
\frac{x+1}{x-y} \left[ y+(y-1) \ln (y+1)+2 \ln ^2(y+1)+\text{Li}_2(-y)+ \frac{\pi ^2}{6}\right]+ (x \leftrightarrow y), \\
I_1^{(2)} &= \frac{x+1}{x-y} 
\Big[
	-y
	+(1-y) \ln (y+1)
	+(1-y) \ln ^2(y+1)
	+\ln (y) \ln ^2(y+1) \\
	&
	-y \text{Li}_2(-y)
	-\frac{5}{3} \ln ^3(y+1)
	+2 \text{Li}_3\left(\frac{1}{y+1}\right)
	- y\frac{\pi ^2}{6}
	-2 \zeta_3
\Big]+ (x \leftrightarrow y).
\ens
To compute  $I_2$, we expand the integral in Eq.~(\ref{eq3.5}) in powers of $\nu$ and obtain the following expression
\bnq
I_2 = I_2^{(0)} + I_2^{(1)} \, \nu + I_2^{(2)} \, \nu^2 + \mathcal{O}(\nu^3),
\enq
with
\begin{align}
I_2^{(0)} &= -\int \limits_{0}^{1} {\rm d}t\;  \chi(t) \; \left(\frac{1}{r_2 r_{12}^{2}}+\frac{1}{r_2^{2} r_{12}}\right)  , \nn
I_2^{(1)} &= \int \limits_{0}^{1} {\rm d}t \; \chi(t) \; \left(
\frac{r_1}{r_2^{2} r_{12}^{2}}
+\frac{\ln \left(r_2\right)}{r_1 r_{12}^{2}}
+\frac{\left(r_{12}^2-2r_2^2\right) \ln \left(r_{12}\right)}{r_1 r_2^{2} r_{12}^{2}}\right) , \\
I_2^{(2)} &= \int \limits_{0}^{1} {\rm d}t \; \chi(t) \; \Bigg(
   -\frac{\left(6+\pi ^2\right) r_1+2 \pi ^2 r_2}{6 r_2^{2} r_{12}^{2}}
   -\frac{\left(r_1-r_2\right) \ln \left(r_{12}\right)}{r_2^{2} r_{12}^{2}}
   -\frac{\ln \left(r_2\right)}{r_2 r_{12}^{2}} \nn
   & \quad 
   -\frac{\left(r_{12}^2-3r_2^2\right) \ln ^2\left(r_{12}\right)}{2 r_1 r_2^{2} r_{12}^{2}}
   -\frac{\ln \left(r_2\right) \ln \left(r_{12}\right)}{r_1 r_{12}^{2}}
   +\frac{\left(r_1+2 r_2\right) \text{Li}_2\left(\frac{r_1}{r_{12}}\right)}{r_2^{2} r_{12}^{2}}
   \Bigg), \nonumber
   \label{eq3.5}
\end{align}
and $\chi(t)$ defined as 
\bnq
\chi = \frac{(x+1)(y+1)}{x-y}\left(\frac{{\rm d} r_1}{{\rm d}t} r_2 - r_1 \frac{{\rm d} r_2}{{\rm d}t}\right).
\enq

We note that   two quadratic polynomials $r_{1,2}$ appear in the denominators  in  
the above integrands. To continue with the $t$-integration, it is important to factorize them.
We find 
\bns
r_1 &= z (t - u)(t - v),
\ens
where $u, v$  are given by the following expression
\bnq
u, v = \frac{y-x+z \pm \sqrt{(z-x-y)^2-4xy}}{2z}.
\enq
In the physical region $(z-x-y)^2-4xy < 0$ so that $u$ and $v$ are complex conjugates of each other.  It is easy to
see that under the transformation $x \leftrightarrow y$ 
\bnq
u \rightarrow 1-v, \quad 
v \rightarrow 1-u.
\enq

For $r_2$, we find
\bns
r_2 &= z (t - t_1 ) (t_2 - t),
\ens
where 
\bnq
t_{\pm} = \frac{1}{2} \left ( 1 \pm \sqrt{1+ \frac{4}{z} } \right ).
\enq
It follows that the two roots $t_{\pm}$ are outside of the integration region $t \in [0,1]$. 

Repeatedly applying partial fractioning 
and integration  by parts, we recognize that all integrals that appear in Eq.~(\ref{eq3.5}) 
are of the following form
\bnq
\int \limits \limits_{0}^{1} \frac{d t}{t-t_a} \; \{1, \ln(t-t_b), \ln(t-t_b)^2, \ln(t-t_c) \ln (t-t_d) \},
\label{eq3.15}
\enq
where  $t_{a,..,d}$ are elements of the following set 
\bnq
\left\{u, v, t_{\pm}, \frac{y+1}{x-y}\right\}.
\enq
Such integrals can be easily written as linear combinations of polylogarithmic functions through weight 3. 

We emphasize one more time that, although the integration over $t$ is straightforward, writing the result in
an optimal way requires some effort.  To this end, we 
will use the variables $x, y, z$ where possible and only employ $u$ and $v$ where
necessary. We will also make the $x \leftrightarrow y$ symmetry  manifest.
Finally, it turns out to be convenient to introduce a new variable $w$ defined as
\be
w = 1 - \frac{1}{t_+}  = 1 + \frac{z}{2} \left ( 1 - \sqrt{1 + \frac{4}{z} } \right ), 
\ee
and to use it when writing the result.  We find 
\bns
C_1^{(0)} &= \frac{(1-w) (x+1) (y+1) (x+y-z-2)}{2 (w+1) \left((x+1) (y+1) z-(x-y)^2\right)}\ln (w)  \\
& \qquad + \frac{(y+1)(x (-x+y+z+1)-y+z)}{(x-y)^2-(x+1) (y+1) z} \ln (x+1) 
+ (x \leftrightarrow y),
\ens
which is explicitly symmetric under $x \leftrightarrow y$. 
For $C_1^{(1)}$, we find 
\bns
C_1^{(1)} = \frac{\pi^2}{6}
+\Big\{\left( A_1 + [B_1 + (u \leftrightarrow v)] \right)
+\left( x \leftrightarrow y, u \rightarrow 1-v, v \rightarrow 1-u \right)
\Big\},
\ens
where
\begin{align}
A_1&= 
	\frac{(x+1)(y+1) \left(w^2+w (y-x)-1\right) }{(w+1) (x-y) (w(y+1)-(x+1))}
	\Bigg[
	\text{Li}_2\left(\frac{(w-1) (x+1)}{w(x+1)-(y+1)}\right) \nn
	& -\text{Li}_2\left(\frac{(w-1) (y+1)}{w(x+1)-(y+1)}\right)
	+\ln \left(\frac{x+1}{y+1}\right) \ln \left(\frac{x-y}{w (x+1)-(y+1))}\right)
	\Bigg] \nn
	&+\frac{y+1}{y-x} \text{Li}_2(-x)
	-\frac{(y+1) (x (-x+y+z+1)-y+z)}{(x-y)^2-(x+1) (y+1) z} \ln ^2(x+1) \nn
	& +\frac{(x+1) (y+1) (w(-x+y+z+2)-2)}{(w+1) (x-y) ((x+1)-w (y+1))}\ln (w) \ln (x+1)  \nn
	& -\frac{(x+1)\left((x-y)^2-z (x+y+2)\right)}{(x-y)^3-(x+1) (y+1) z(x-y)} \ln (w) \ln ((1-w) (y+1)), \\
B_1 &= 
\frac{(x+1) (y+1) }{(1-(u-1) u z) (x-y)}
\Bigg[\text{Li}_2\left(\frac{1}{u (1-w)+w}\right)-\text{Li}_2\left(\frac{w}{u (1-w)+w}\right) \nn
   & -\text{Li}_2\left(\frac{x+1}{u(x-y)+y+1}\right)+\frac{1}{2} \ln (w) \ln \left(\frac{u-1}{u}\right) \nn
   & -\ln (w) \ln
   \left(\frac{u}{u (1-w)+w}\right)+\ln (y+1) \ln \left(\frac{u (x-y)}{u(x-y)+y+1}\right)
   \Bigg]. \nonumber
\end{align}
It is easy to check that $A_1$ is real and that 
\bnq
B_1 \big|_{u \leftrightarrow v} = B_1^{*}.
\enq
Hence, the final result for $C_1^{(1)}$ reads 
\bns
C_1^{(1)} = \frac{\pi^2}{6}
+\Big\{
 A_1 + 2\text{Re}[B_1] 
+\left( x \leftrightarrow y, u \rightarrow 1-v, v \rightarrow 1-u \right)
\Big\}.
\ens

The function  $C_1^{(2)}$ admits  a  representation  similar to  $C_1^{(1)}$
\bns
C_1^{(2)} &= -2\zeta_3
+\left( A_{2,1} + 2\text{Re}[A_{2,2}] \right) \\
& \qquad +\Big\{ B_{2,1} + 2\text{Re}[B_{2,2}] 
+\left( x \leftrightarrow y, u \rightarrow 1-v, v \rightarrow 1-u \right)
\Big\}.
\ens
The functions $A_{2,1}, A_{2,2}, B_{2,1}, B_{2,2}$  are, unfortunately,
very lengthy and we do not  present them here. However, they 
can be found in the ancillary file 
provided with this submission.

\section{Numerical results}
\label{sect:numres}

We now discuss the numerical results of the calculation.  We consider proton collisions at the LHC running at an energy of $13.6~{\rm TeV}$ and adopt parameters and selection criteria from Ref.~\cite{Asteriadis:2023nyl}. 
The Higgs boson is considered stable with a mass of $m_H =125~{\rm GeV}$.
The vector boson masses are set to $m_W= 80.398~{\rm GeV}$ and $m_Z = 91.1876~{\rm GeV}$ with widths of $\Gamma_W = 2.105~{\rm GeV}$ and $\Gamma_Z = 2.4952~{\rm GeV}$ respectively.
The Fermi constant value of $G_F = 1.16639 \times 10^{-5}~{\rm GeV}^{-2}$ is used to derive the weak couplings and the CKM matrix is set equal to the identity matrix.

We use \texttt{NNPDF31-nnlo-as-118} parton distribution
functions~\cite{NNPDF:2017mvq} and $\alpha_s(m_Z) = 0.118$ for all
calculations reported below. The evolution of both parton distribution functions
and the strong coupling constant is obtained directly from
LHAPDF~\cite{Buckley:2014ana}.
We fix the factorization scale $\mu_F = m_H$ throughout this calculation.

We require that a WBF  event contains at least two jets with transverse momenta higher than
$25~{\rm GeV}$. The tagging jets are required to have  rapidities between $-4.5 < y < 4.5$ and should be  separated by a large rapidity interval $|y_{j_1} - y_{j_2}| > 4.5$.
Their invariant mass should be larger than $600~{\rm GeV}$. The two jets 
should be in opposite hemispheres in the laboratory frame; this is enforced by requiring
the product of their rapidities to be negative, $y_{j_1} y_{j_2} < 0$.
In principle, jet identification requires a particular jet algorithm but this is not relevant with only two quarks in the final state.  For the results  reported below, we employ the two-loop
amplitude in the leading eikonal approximation as summarized in
Eq.~(\ref{eq2.8}),  and we do not include the next-to-eikonal power
correction computed in Ref.~\cite{Long:2023mvc}.

The main conclusion of our analysis is that the  ${\cal O}(\beta_0 \alpha_s^3)$ corrections  significantly reduce
the scale dependence of the  non-factorizable contributions.  To illustrate this point,
we first show results for fiducial \emph{non-factorizable} contributions to the WBF
cross section at leading ${\cal O}(\alpha_s^2)$ order, and then
compare them  with the results at
next-to-leading order where  we only include the ${\cal O}(\beta_0 \alpha_s^3)$
correction. We find
\be
\sigma^{\rm LO}_{\rm nf} = -2.97^{-0.69}_{+0.52}~{\rm fb},\;\;\;\; \sigma^{\rm NLO}_{\rm nf} = -3.20^{-0.01}_{+0.14}~{\rm fb},
\label{eq3.1}
\ee
where we have used the renormalization scale $\mu = m_H$ to obtain the central  values,
and $\mu = m_H/2$ and $\mu = 2 m_H$ to obtain values described by superscripts and subscripts in Eq.~(\ref{eq3.1}),
respectively.  It follows
from Eq.~(\ref{eq3.1})  that the scale variation is reduced very significantly once ${\cal O}(\beta_0 \alpha_s^3)$
contributions are included.

\begin{figure}[t]
\centering
\includegraphics[width=0.45 \textwidth]{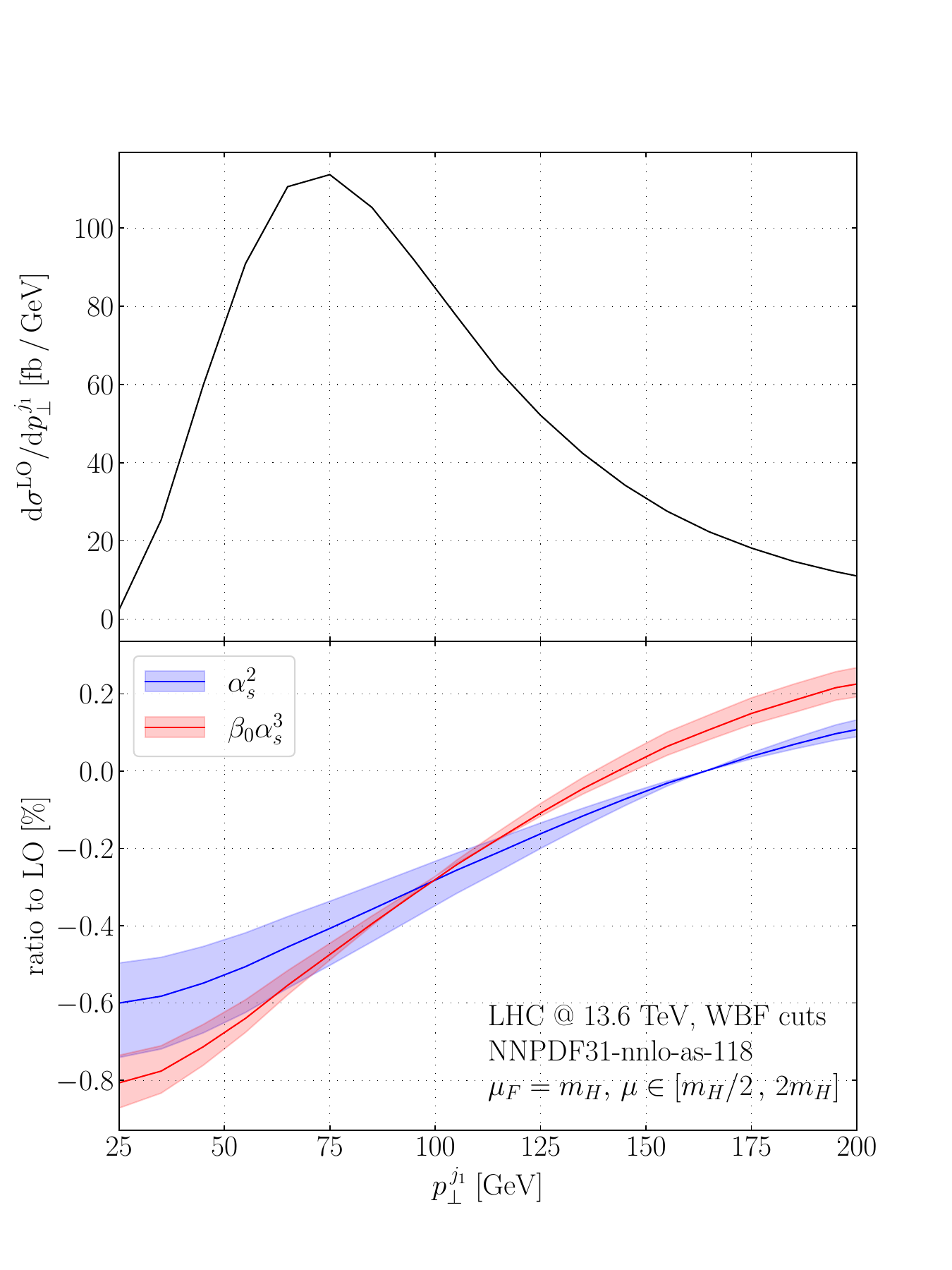} \includegraphics[width=0.45 \textwidth]{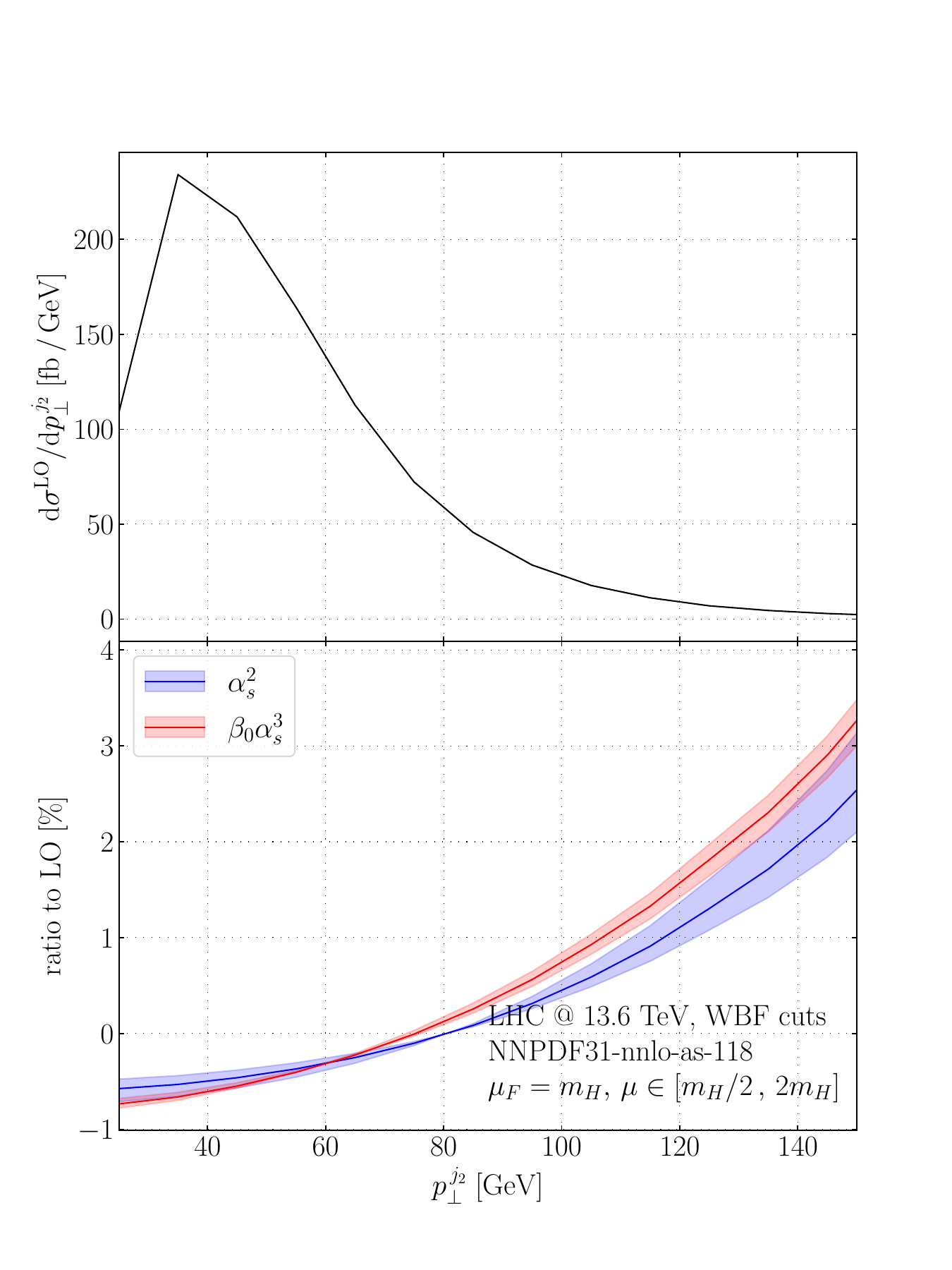}
\caption{Transverse momenta distributions of hardest and next-to-hardest jets in Higgs boson production in
  weak boson fusion.
	The upper panes show the LO (tree-level) distributions and the lower panes show the ratio of non-factorizable contributions to LO for corrections of ${\cal O}{(\alpha_s^2)}$ (blue) and ${\cal O}{(\beta_0 \alpha_s^3)}$ (red).
The factorization scale, $\mu_F$, is kept fixed and only the renormalization scale, $\mu$, is varied.
See text for further details.}
\label{fig::ptjs}
\end{figure}

\begin{figure}[t]
\centering
\includegraphics[width=0.45 \textwidth]{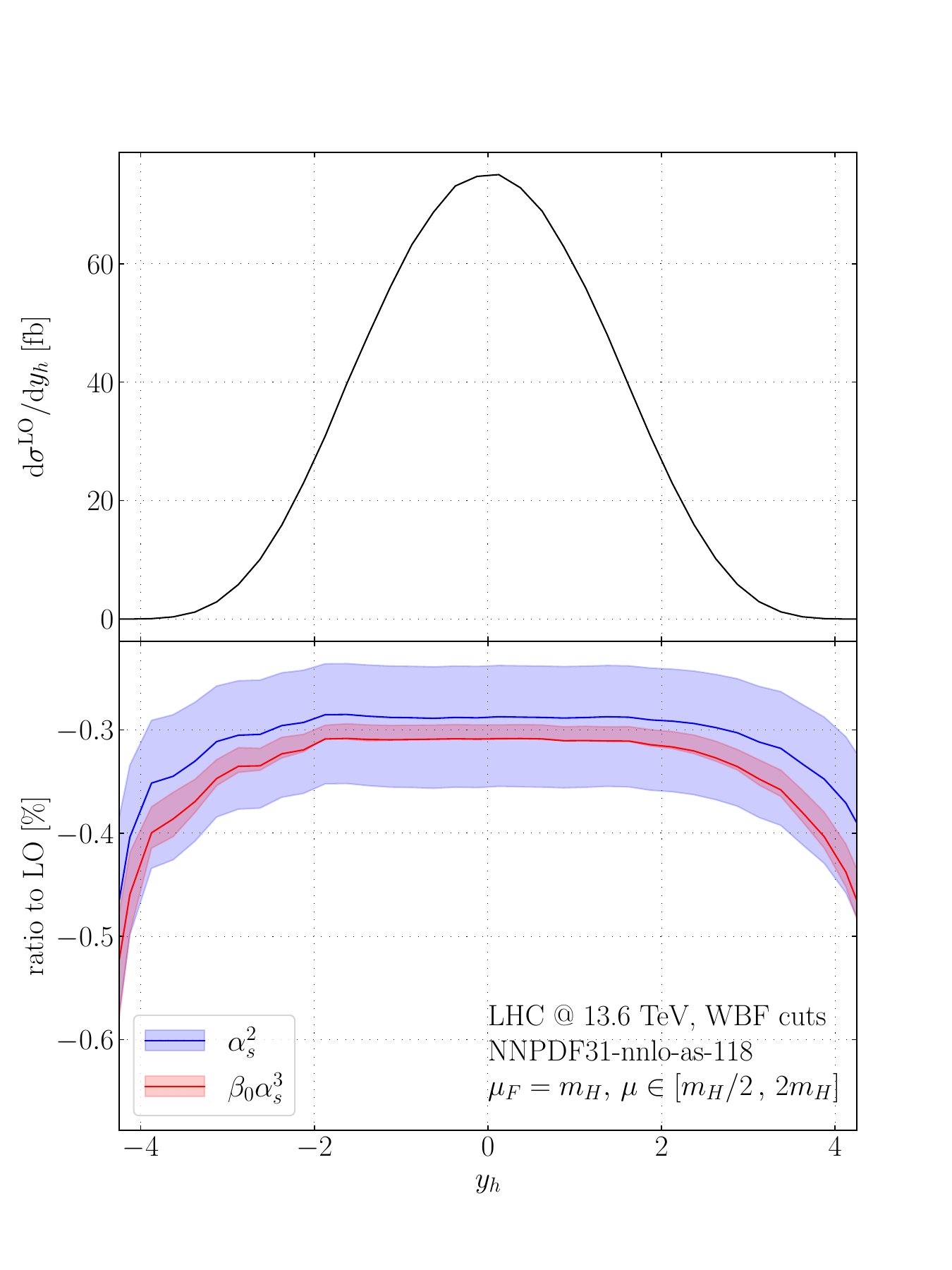} \includegraphics[width=0.45 \textwidth]{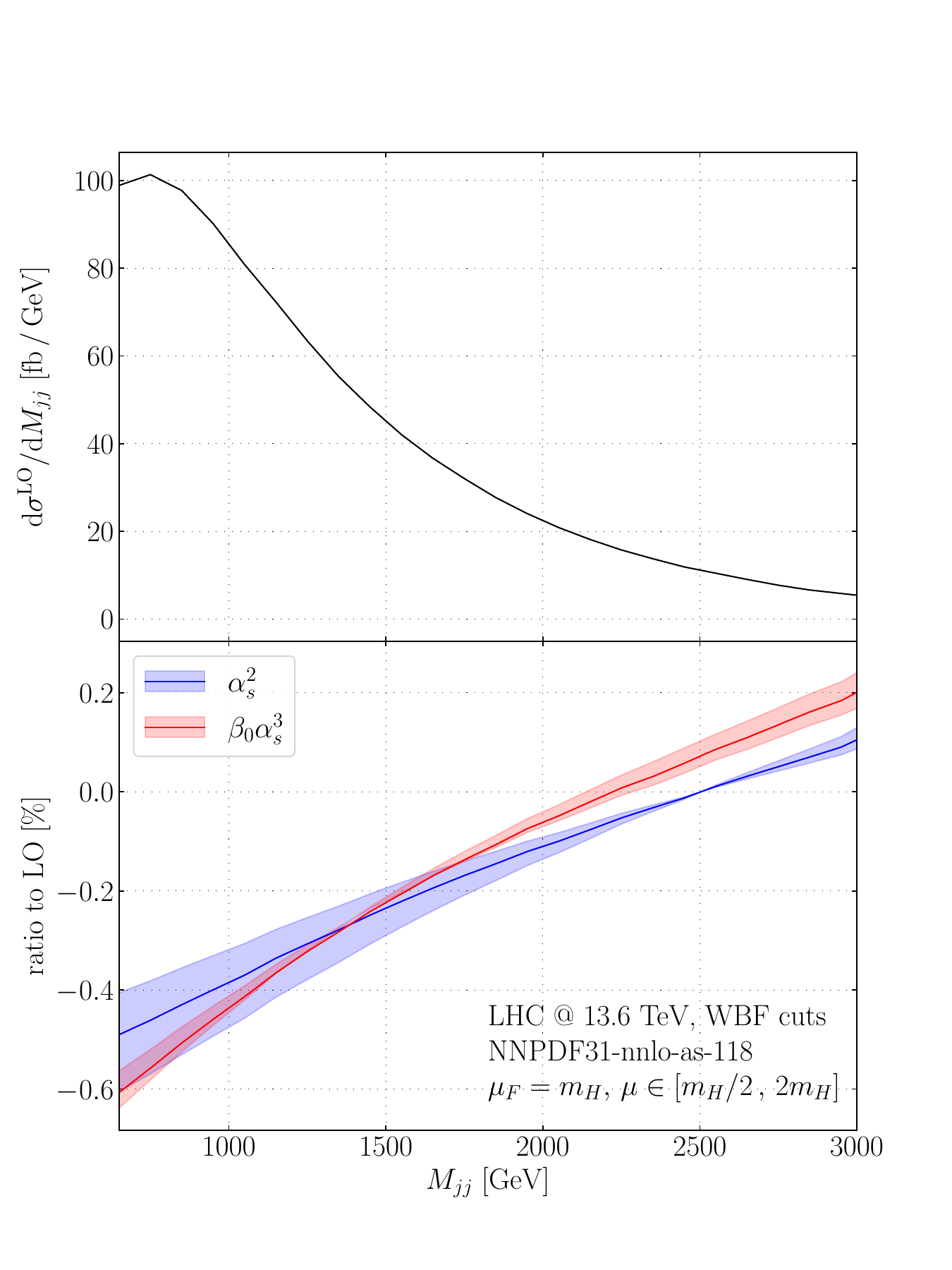}
\caption{Non-factorizable corrections to
  Higgs boson rapidity distribution (left panes)  and the invariant mass distribution of the two final state jets (right panes).
	The upper panes show the LO (tree-level) distributions and the lower panes show the ratio of non-factorizable contributions to LO for corrections of ${\cal O}{(\alpha_s^2)}$ (blue) and ${\cal O}{(\beta_0 \alpha_s^3)}$ (red).
The factorization scale, $\mu_F$, is kept fixed and only the renormalization scale, $\mu$, is varied.
See text for further details.}
\label{fig::rapinvmass}
\end{figure}

The same statement applies to kinematic distributions.  We illustrate this in Figures~\ref{fig::ptjs} and \ref{fig::rapinvmass} where examples of
transverse momenta, rapidity and two-jet invariant mass distributions are shown.  In all plots,  the upper
pane displays  the leading order (tree-level) distribution.
In the lower pane ratios of non-factorizable corrections to leading order  distributions are shown. Bands correspond
to the range of theoretical predictions obtained  with the renormalization scales from  the interval $m_H/2 \le \mu \le 2 m_H$
at leading ${\cal O}(\alpha_s^2)$ and next-to-leading ${\cal O}(\beta_0 \alpha_s^3)$ orders, respectively. 
We observe that accounting for NLO ${\cal O}(\beta_0 \alpha_s^3)$
corrections stabilizes theoretical predictions
by strongly reducing their dependence on the renormalization scale.   We note that
the ${\cal O}(\beta_0 \alpha_s^3)$  results are sometimes outside  the ${\cal O}(\alpha_s^2)$
scale-variation bands; this mostly happens at large(r)  values of the transverse momenta and
invariant masses. 

\section{Conclusions}
\label{sect:conclusion}
In this paper we computed ${\cal O}(\beta_0 \alpha_s^3)$ corrections to the non-factorizable contribution to Higgs
boson production in weak boson fusion.  These corrections  reflect the impact  of the running of the QCD  coupling
constant on the non-factorizable contribution to the WBF
cross section and are responsible for stabilizing the dependence of the theoretical prediction 
on the renormalization scale.  Indeed, we find that after including these ${\cal O}(\beta_0 \alpha_s^3)$
corrections, the dependence of the cross section on the renormalization scale reduces from about ${\cal O}(20)$ percent
to below ${\cal O}(5)$ percent. Similar reductions of the scale dependence are observed in theoretical predictions
for major kinematic distributions including transverse momenta and rapidity distributions
of the tagging jets and the Higgs boson. 

We provided a simple one-dimensional integral representation of the ${\cal O}(\beta_0\alpha_s^3)$ non-factorizable
corrections  as well as the analytic formulas for these corrections.  Although the analytic results 
are  complex, they can easily be implemented into partonic Monte Carlo and used to obtain phenomenological predictions.
In fact, we have used the one-dimensional integral representation of these corrections to cross check the results of the
analytic computation.  Under realistic running conditions, analytic formulas provide a significant
speed-up whereas a one-dimensional integral representation is a  slow but robust way to compute cross sections and distributions. 

\vspace*{0.3cm}
\noindent{\bf Acknowledgments:}
This research of K.M. and M.M.L is  partially supported  by the Deutsche Forschungsgemeinschaft (DFG, German Research Foundation)
under grant 396021762 - TRR 257.
The work of C.B.H. presented here is supported by the Carlsberg Foundation, grant CF21-0486.
K.M. would like to thank the Galileo Galilei Institute for Theoretical Physics in Florence for the hospitality and the INFN for partial support during the completion of this work.
    
\bibliographystyle{JHEP}
\bibliography{vbfvirtbub}

\end{document}